\documentclass[
aps,%
12pt,%
final,%
notitlepage,%
oneside,%
onecolumn,%
nobibnotes,%
nofootinbib,%
superscriptaddress,%
noshowpacs,%
centertags]%
{revtex4}
\newcommand{\bc}[2]{\langle {#1} , {#2} \rangle}
\newcommand{\Mnm}{\mbox{M}(2^n ,m;{\bf C})}
\newcommand{\Mnmone}{\mbox{M}_{1}(2^n ,m;{\bf C})}
\newcommand{\trace}{\mbox{{\rm tr}} \;}
\newcommand{\Herpos}{H^{m}_{+, \, 1}}
\newcommand{\Herposemi}{H^{m}_{\geq 0, \, 1}}
\newcommand{\Rmet}[2]{(\!( {#1} , {#2} )\!)^{R}}
\newcommand{\QFmet}[2]{(\!( {#1} , {#2} )\!)^{QF}}
\newcommand{\inv}{\pi_m^{-1}(\Herpos)}
\newtheorem{definition}{Definition}
\newtheorem{theorem}[definition]{Theorem}

\newtheorem{proposition}[definition]{Proposition}

\begin{document}

\title{
Certain integrable system on a space
\\
associated with a quantum search algorithm
}
\author{\firstname{Y.}~\surname{Uwano}}
\email{uwano@fun.ac.jp}
\altaffiliation[\, Previous address (till to March 2005):]
{%
\, Department of Applied
\\ \vskip -30pt
Mathematics and Physics, Kyoto University, Kyoto 606-8501,
Japan
}%
\thanks{${}$
Supported by Grant-in-Aid
\\ \vskip -30pt
Scientific Research No.16560060 from JSPS.
}
\affiliation
{%
School of Systems Information Sciences, Future University-Hakodate,
\\
116-2 Kamedanakano-cho, Hakodate, Hokkaido 041-8655, Japan
}%
\author{\firstname{H.}~\surname{Hino}}
\altaffiliation[Present address:]
{%
\, Security Technology Research Center,
Systems Development Laboratory,
Hitachi, Ltd., 
\\ \vskip -30pt
1099 Ohzenji, Asao-ku, Kawasaki 215-0013, Japan.
}%
\author{\firstname{Y.}~\surname{Ishiwatari}}
\altaffiliation[Present address:]
{%
\, Solution Services No.2, Service Delivery-Communications,
IBM Global Services Japan,
\\ \vskip -30pt
19-21 Nihonbashi, Hakozaki-cho, Chuo-ku, Tokyo 103-8510, Japan.
}%
\affiliation{%
Department of Applied Mathematics and Physics,
Kyoto University, Kyoto 606-8501, Japan
}%
\begin{abstract}
On thinking of a Grover-type quantum search algorithm for
an ordered tuple of multi-qubit states, a gradient system
associated with the negative von-Neumann entropy is studied
on the space of regular relative-configurations of multi-qubit
states (SR${}^2$CMQ). The SR${}^2$CMQ emerges, through
a geometric procedure, from the space of ordered tuples of
multi-qubit states for the quantum search. The aim of this
paper is to give a brief report on the integrability of
the gradient dynamical system together with quantum information
geometry of the underlying space, SR${}^2$CMQ, of that system.
\end{abstract}

\maketitle

\section{Introduction}
Quantum computing has been investigated as one of the most
challenging research subjects \cite{NC} over a decade. 
In 2001, Miyake and Wadati \cite{MW} provided a differential
geometric characterization of the Grover search algorithm
for a single target state \cite{G} from a geometric
viewpoint: They applied the fiber bundle structure of the
space of normalized multi-qubit states over the complex
projective space to the sequence of states generated by
the search algorithm. The sequence after projection is shown
to be along a geodesic of the complex projective space
endowed with the Fubini-Study metric.
\par
The aim of this paper is to report very briefly on a
integrable dynamical system arising from geometric studies
on a quantum search for an ordered tuple of multi-qubit
states, which possesses rich quantum information features:
The space for the integrable system is thought of as a
\lq quantum information space' since it is represented
mathematically as the space of positive definite Hermitean
matrices with unit trace ({\it i.e.}, regular density
matrices) endowed with the symmetric logarithmic derivative
(SLD) quantum Fisher metric. Further, the integrable system
to be dealt with is the gradient system with the negative
von-Neumann entropy as the potential. The quantum
information space structure presented in this paper emerges
in a purely geometric way motived by Miyake and Wadati
\cite{MW,Rm1}. The quantum information space structure with
the SLD quantum Fisher metric is shown to be equivalent to
the Riemannian structure arising from the fibered space
structure over the space of regular \lq relative
configurations' of multi-qubit states in ordered tuples.
The organization of this paper is outlined in what
follows.
\par
Section~II is a preliminary section. The Hilbert space for a
Grover-type search algorithm for an ordered tuple of
multi-qubit states is presented together with a brief
description of the search algorithm. The space of normalized
ordered tuples of multi-qubit states where the search is
proceeded will be abbreviated to STMQ.
\par
In Section~III, the space of regular
\lq relative-configurations' of multi-qubit states in ordered,
tuples abbreviated to SR${}^2$CMQ, is studied from geometric
viewpoint. As the first step, the quotient space of STMQ
under the left $U(2^n)$ action is introduced as the space of
\lq relative-configurations' of multi-qubit states in ordered
tuple, abbreviated to SRCMQ. As the regular part of the
SRCMQ free from singularities, SR${}^2$CMQ is introduced
and is shown to be isomorphic to the space of positive
definite Hermitean matrices with unit trace. A pair of
geometric structures are introduced to SR${}^2$MQ:
One is the Riemannian metric arising from the fibered
space structure over SR${}^2$CMQ. The other is the SLD
(symmetric logarithmic derivative) quantum Fisher metric,
on looking SR${}^2$MQ mathematically upon as the space of
non-singular density matrices. As one of the goals of this
paper, the SLD quantum Fisher metric is shown to be identical
to the Riemannian metric given above up to a multiplication
constant.
\par
Since SR${}^2$MQ is shown to be isomorphic to the space of
non-singular density matrices endowed with the SLD quantum
Fisher metric in Sec.~III and since the von-Neumann entropy
is known to be a typical quantum-information object on
SR${}^2$MQ, it would be very interesting to study the
gradient system associated with the von-Neumann entropy
in Section~IV.
On using a geometric calculus, the equation of motion for
the gradient system is solved, so that the gradient system
is shown to be integrable.
\par
Section~V is for concluding remarks. A pair of papers
\cite{UIH, UHI} are in preparation which deal with
the geometric study on SR${}^2$MQ and the gradient system on
SR${}^2$MQ in more detail.
\section{Preliminaries}
As is known very well, the Hilbert space for multi-qubit
systems is the tensor product,
$
\overbrace{({\bf C}^2) \otimes \cdots \otimes ({\bf C}^2)}^{n}
\cong ({\bf C}^{2})^{\otimes n} ,
$
of the single-qubit space, ${\bf C}^2$ \cite{NC}.
We prepare the direct sum,
\begin{equation}
\label{m-sum}
\overbrace{({\bf C}^2)^{\otimes n} \oplus \cdots \oplus
	           ({\bf C}^2)^{\otimes n} }^{m} ,
\end{equation}
of the Hilbert space $({\bf C}^2)^{\otimes n}$ for $n$-qubit
states to describe ordered tuples of multi-qubit states,
so that any ordered tuple of $n$-qubit states is expressed
as
\begin{equation}
\label{Phi}
\Phi = ( \phi^{(1)}, \, \cdots , \, \phi^{(m)} )
\qquad
(\phi^{(j)} \in ({\bf C}^2)^{\otimes n}, \; j=1, \cdots , m) ,
\end{equation}
where $m$ indicates the number of $n$-qubit states in
ordered tuples.
On writing out each $\phi^{(j)} \in ({\bf C}^2)^{\otimes n}$
in $\Phi$ of (\ref{Phi}) to be the column-vector form,
\begin{equation}
\label{phi^(j)}
\phi^{(j)}
=
(\phi^{(j)}_{1} , \, \phi^{(j)}_{2}, \, \cdots ,
\phi^{(j)}_{2^n} )^T \in ({\bf C}^2)^{\otimes n}
\quad
(j=1,2, \cdots , m; \, \, {}^T \, \mbox{the transpose}),
\end{equation}
$\Phi$ of (\ref{Phi}) has the complex $2^n \times m$ matrix
form with the components
\begin{equation}
\label{Phi-mat-rep}
\Phi_{h\ell}
=
\phi^{(\ell)}_{h}
\qquad
(h=0,\cdots, 2^n-1; \, \ell =1,\cdots ,m).
\end{equation}
We can thereby identify the space of ordered tuples of
$n$-qubit states with the space of complex $2^n \times m$
matrices denoted by $\Mnm$ henceforth. The $\Mnm$ admits
the conventional Hilbert-space structure associated with
the Hermitean inner product,
\begin{eqnarray}
\label{inner}
\langle \Phi , \, \Phi^{\prime} \rangle
=
\frac{1}{m} \trace (\Phi^{\dagger} \Phi ) 
\qquad
(\Phi , \Phi^{\prime} \in \Mnm).
\end{eqnarray}
As the computational basis, the set of matrices,
$\Phi (j;k)$ ($j=1, \cdots , m$, $k=0, \cdots , 2^n -1$),
with the components
\begin{eqnarray}
\label{ON-multi}
&&
(\Phi (j ; k))_{h \ell}
=
\sqrt{m} \, \delta_{jh} \delta_{k\ell}
\qquad
(h=0, \cdots , 2^n -1; \, \ell=1, \cdots , m) 
\end{eqnarray}
is taken, where $\delta_{jh}$ and $\delta_{k\ell}$ indicate
the Kronecker delta. The subset,
\begin{equation}
\label{Mnmone}
\Mnmone
=
\left\{ \Phi \in \Mnm \, \left\vert \,
\bc{\Phi}{\Phi}
= \frac{1}{m} \trace (\Phi^{\dagger} \Phi ) 
=1 \right. \right\} ,
\end{equation}
of $\Mnm$ is taken as the space of \lq quantum states',
which is diffeomorphic to the unit sphere of dimension
$2^{n+1}m-1$. The space $\Mnmone$ represents the space of
normalized ordered tuples of multi-qubit states, STMQ.
\par
The Grover-type search algorithm for an ordered tuple of
multi-qubit states is described in the following. Let us
start with defining the initial tuple to be the sum of all
the computational-base states with equal weight;
\begin{equation}
\label{A}
A = 
\frac{1}{\sqrt{2^n m}} \sum_{j=1}^{m} \sum_{k=0}^{2^n-1}
\Phi ( j ; \, k ) \, \in \Mnmone.
\end{equation}
The target tuple denoted by $W \in \Mnmone$ is assumed to
be the tuple of distinct computational base-states;
\begin{eqnarray}
\label{W}
W= \frac{1}{\sqrt{m}}
( w^{(1)} , w^{(2)} , \cdots , w^{(m)}) ,
\quad
w^{(h)} 
= \Phi ( {}^{\exists} j_{h} ; \, {}^{\exists}k_{h} )  
\quad \mbox{with} \quad
w^{(h)} \ne w^{(h^{\prime})}  \quad (h \ne h^{\prime}).
\end{eqnarray}
Like the original case due to Grover \cite{G},
the search process is made by applying the composition
$-I_A \circ I_W$ of unitary operators,
\begin{equation}
\label{IW-IA}
I_W (\Phi) = \Phi - 2 \bc{W}{\Phi} W 
\quad \mbox{and} \quad
I_A (\Phi) = \Phi - 2 \bc{A}{\Phi} A
\quad (\Phi \in \Mnm) ,
\end{equation}
repeatedly to the initial tuple $A$. By calculation,
we have
\begin{eqnarray}
\label{iteration}
(-I_A \circ I_W)^k (A) 
=
\Big( \cos (k+\frac{1}{2}) \theta \Big) \, R
	+ \Big( \sin (k+\frac{1}{2}) \theta \Big) \, W
\qquad (k=0,1,2,\cdots)
\end{eqnarray}
with $R= \sqrt{2^n /(2^n-1)}A - \sqrt{1/(2^n-1)}W$ and
$\sin (\theta/2) = \sqrt{1/2^n}$ ($\theta \in (0,\pi))$.
In the case that the degree of qubit, $n$, is large enough,
the probability of finding $W$ in
$(-I_A \circ I_W)^{k}A$
gets closed to one as $k$ does to $(\pi \sqrt{2^n}-1)/2$.
\section{The geometry of SR${}^2$MQ}
In this section, we study the space, SR${}^2$CMQ, of
regular \lq relative-configurations' of multi-qubit states
in ordered tuples both from the fibered space structure and
the quantum-information structure viewpoints.
\subsection{Setting-up: SRCMQ and SR${}^2$CMQ}
We start with giving a description of the
relative-configuration of multi-qubit states in ordered tuples.
Let us take $\Phi \in \Mnmone$ all of whose column vectors,
$\phi^{(j)}$s, in (\ref{Phi}) are non-vanishing \cite{Rm2}.
By the placement of the column vectors $\phi^{(k)}$ ($k > 1$)
relative to $\phi^{(1)}$, we mean the relative-configuration
of multi-qubit states in that ordered tuple $\Phi$.
For $\Phi \in \Mnmone$ with non-vanishing columns, let us
consider the matrix $g\Phi \in \Mnmone$ with $g \in U(2^n)$,
each of whose column vectors, $g \phi^{(j)}$s, are given by
the unitary action of $g$ to $\phi^{(j)}$s by $g$.
Then we can say that the relative-configurations of
multi-qubit states in $\Phi$ and those in $g \Phi$ are the
same. Accordingly, as the space, SRCMQ, of 
relative-configurations of multi-qubit states, the quotient
space $\Mnmone /\! \sim$ of $\Mnmone$ is thought of, which is
associated with the equivalence relation
\begin{equation}
\label{U-equiv}
\Phi \sim \Phi^{\prime}
\qquad \mbox{if and only if} \qquad
{}^{\exists} g \in U(2^n) \quad 
\mbox{s.t.} \quad \Phi^{\prime}=g \Phi ,
\end{equation}
generated by the $U(2^n)$ action on $\Mnmone$.
For $\Mnmone / \! \sim$, we have the following.
\begin{proposition}
\label{quotient}
The quotient space $\Mnmone / \! \sim$ is isomorphic
the space of positive semi-definite $m \times m$
Hermitean matrices with unit trace,
\begin{eqnarray}
\label{Her-pos}
\Herposemi 
=
\{ \rho \in M(m,m;{\bf C}) \, \vert \, 
\rho^{\dag}=\rho , \, \trace \rho =1, \,
\rho : \mbox{positive semi-definite} \} .
\end{eqnarray}
The space, SRCMQ, of relative configurations of multi-qubit
states in ordered tuples will be represented in the form
$\Herposemi$, henceforth. 
\end{proposition}
The proof is accomplished by showing that the map,
\begin{equation}
\label{pi_m}
\pi_m : \Phi \in \Mnmone 
\mapsto \frac{1}{m} \Phi^{\dag} \, \Phi \in \Herposemi ,
\end{equation}
is surjective and that $\pi_m (\Phi)=\pi_m(\Phi^{\prime})$
holds true iff so does (\ref{U-equiv}) \cite{Rm3}.
\par
We wish to specify a \lq regular' part of SRCMQ on which
we can make differential calculus without any problems.
In view of (\ref{Her-pos}), $\Herposemi$ (SRCMQ) has a
natural boundary,
$
\partial \Herposemi
=
\{ \rho \in \Herposemi \, \vert \, \det \rho =0 \}
$
,
so that we have
\begin{eqnarray}
\label{remove-bd}
\Herpos := \Herposemi - \partial \Herposemi
=
\{ \rho \in M(m,m;{\bf C}) \, \vert \, \rho^{\dagger}=\rho,
 \, \det \rho >0, \, \trace \rho =1 \} .
\end{eqnarray}
The $\Herpos$ is nothing but the set of positive definite
$m \times m$ Hermitean matrices with unit trace. We have
the following proposition providing a differentiable
structure to $\Herpos$.
\begin{proposition}
\label{fiber-SRCMQ}
The inverse image, $\inv$ of $\Herpos$ by $\pi_m$ is made
into the fiber bundle, $\pi_m : \inv \rightarrow \Herpos$,
with the fiber diffeomorphic to $U(2^n)/U(2^n-m)$.
\end{proposition}
We here present only a key to the proof: A key is that
the isotropy subgroup,
$
G_{\Phi}
=
\{ g \in U(2^n) \, \vert \, L_g(\Phi) = g\Phi =\Phi \}
$
,
of $U(2^n)$ at any $\Phi \in \pi_m^{-1}(\Herpos)$ is
isomorphic to $U(2^n-m)$ \cite{UHI}. According to
transformation-group theory \cite{K}, this fact implies that
any $\Phi \in \pi_m^{-1}(\Herpos)$ has the same orbit type
in common. This proves our assertion.
\par
We wish to give an account of referring to $\Herpos$ as
SR${}^2$CMQ. From (\ref{Her-pos}), it is easy to see that
any $\Phi \in \pi_m^{-1}(\Herpos)$ is of rank $m$.
This implies the linear independence among the column
vectors of any $\Phi \in \pi_m^{-1}(\Herpos)$, while the
linear independence does not hold true for any
$\Phi \in \pi_m^{-1}(\partial \Herposemi)$. We may thereby
think any $\Phi \in \pi_m^{-1}(\Herpos)$ is endowed with a
sort of regularity in the relative-configuration.
We reach to the following.
\begin{definition}
\label{def-reg}
The space, SR${}^2$CMQ, of regular relative-configurations
of multi-qubit states in ordered tuples is represented as
the space of positive definite Hermitean matrices with unit
trace $\Herpos$.
\end{definition}
\subsection
{The fibered space structure and the Riemannian metric}
We are to give the Riemannian metric of $\Herpos$, namely
SR${}^2$CMQ, that makes the projection,
$
\pi_m : \pi_m^{-1}(\Herpos) \rightarrow \Herpos
$, 
the Riemannian submersion. We start with the Riemannian
metric of $\inv$. Recalling the fact that $\inv$ is open
in $\Mnmone$ together with the definition (\ref{Mnmone})
of $\Mnmone$, the tangent space of $\inv$ at $\Phi$ is
given by
\begin{eqnarray}
\label{T_Phi-inv}
T_{\Phi} \inv
=
\{ X \in \Mnm \, \vert \,
	\Re (\trace (\Phi^{\dagger}X))=0 \} 
\qquad (\Phi \in \inv) . 
\end{eqnarray}
From the Hermitean inner product (\ref{inner}) of $\Mnm$
($\supset \inv$), the Riemannian metric denoted by
$(\cdot, \cdot)$ arises, which endows the inner product 
\begin{eqnarray}
\label{metric-inv}
(X , X^{\prime} )_{\Phi}
=
\frac{1}{m} \Re ( \trace (X^{\dagger} X^{\prime}))
\qquad (X , X^{\prime} \in T_{\Phi} \inv)
\end{eqnarray}
to any $T_{\Phi} \inv$. Using the metric $(\cdot, \cdot)$
of $\inv$, we wish to give the orthogonal direct-sum
decomposition,
\begin{eqnarray}
\label{o-decomp}
T_{\Phi}\inv
=
\ker (\pi_{m *, \Phi}) 
\oplus (\ker (\pi_{m *, \Phi}))^{\perp},
\end{eqnarray}
explicitly, where $\ker (\pi_{m *, \Phi})$ denotes the
kernel of the tangent map,
\begin{eqnarray}
\label{def-tangent-map}
(\pi_m)_{*,\Phi}:
X \in T_{\Phi} \inv
\mapsto
\frac{1}{m}
\left( \Phi^{\dagger}X + X^{\dagger} \Phi \right)
\in T_{\pi_m(\Phi)} \Herpos 
\quad (\Phi \in \inv),
\end{eqnarray}
of $\pi_m$ at $\Phi$ and  $\ker (\pi_{m *, \Phi})^{\perp}$
the orthogonal complement of the kernel with respect to
$(\cdot, \cdot)_{\Phi}$.
\begin{proposition}
\label{str-ver-hor}
Let a given $\Phi \in \pi_m^{-1}(\Herpos)$ admit the
singular decomposition \cite{RM,Rm3},
\begin{eqnarray}
\nonumber
\Phi 
&=& g 
\left( \begin{array} {c} \sqrt{m} \, 
	\Lambda \\ O_{2^n-m, m} \end{array} \right)
h^{\dagger} 
\quad \mbox{with} \quad
g \in U(2^n), \, h \in U(m) ,
\\
\label{sing-decomp}
&&
\Lambda = \mbox{{\rm diag}} (\lambda_1, \cdots , \lambda_m)
\quad
\mbox{s.t.}
\quad
\sum_{j=1}^m \lambda_j^2 =1, \quad \lambda_j > 0 \, \,
(j=1, \cdots ,m).
\end{eqnarray}
Then
$
\ker (\pi_{m *, \Phi})$ and $\ker (\pi_{m *, \Phi})^{\perp}
$
take the following forms:
\begin{eqnarray}
\nonumber
\ker (\pi_{m *, \Phi})
&=&
\left\{
X \in T_{\Phi}\pi_m^{-1}(\Herpos) \, 
\left\vert
\,
X=g \left(
  	\begin{array}{c}
  	(\sqrt{m}\Lambda)^{-1} \eta \\ x 
  	\end{array} 
  	\right) h , 
\right.
\right.
\\
\label{Ver-exp}
&& \qquad \qquad
\left.
\phantom{
  	\begin{array}{c}
  	(\sqrt{m}\Lambda)^{-1} \eta \\ x 
  	\end{array} 
}
\eta \in u(m), \, x \in M(2^n-m,m;{\bf C})
\right\} ,
\\
\nonumber
\ker (\pi_{m *, \Phi})^{\perp}
&=&
\left\{
X \in T_{\Phi}\pi_m^{-1}(\Herpos) \, 
\left\vert \,
X=g \left(
  	\begin{array}{c} (\sqrt{m}\Lambda)^{-1} 
  	(\sigma + \alpha_{\Lambda}(\sigma) ) 
  	\\
  	O_{2^n-m,m} \end{array} 
  	\right) h , 
\right. \right.
\\
\label{Hor-exp}
&&
\qquad \qquad
\left.
\phantom{
  	\begin{array}{c}
  	(\sqrt{m}\Lambda)^{-1} \sigma \\ O_{2^n-m,m} 
  	\end{array} 
}
\sigma \in M(m,m; {\bf C}), \, \sigma^{\dagger}=\sigma ,
\trace \sigma =0
\right\} ,
\end{eqnarray}
where $M(2^n-m,m;{\bf C})$ and $O_{2^n-m,m}$ denote the
set of $(2^n-m) \times m$ complex matrices and the
$(2^n-m) \times m$ null matrix, respectively.
The $\alpha_{\Lambda}$ in (\ref{Hor-exp}) is the linear
map of $m \times m$ traceless Hermitean matrices into the
anti-Hermitean matrices whish is defined to satisfy
\begin{eqnarray}
\label{alpha}
\Lambda^{-2} \alpha_{\Lambda} (\sigma)
+
\alpha_{\Lambda}(\sigma) \Lambda^{-2}
=
-\Lambda^{-2} \sigma + \sigma \Lambda^{-2} ,
\end{eqnarray}
where $\Lambda$ appears in the singular-value
decomposition (\ref{sing-decomp}).
\end{proposition}
\par
We move on to define the horizontal lift of tangent vectors
of $\Herpos$, where the tangent space $T_{\rho} \Herpos$ at
$\rho \in \Herpos$ is given by the space of traceless
$m \times m$ Hermitean matrices,
\begin{eqnarray}
\label{tan-Herpos}
T_{\rho} \Herpos
=
\{ \Xi \in M(m,m;{\bf C}) \, \vert \, 
\Xi^{\dagger}=\Xi, \, \trace \Xi =0\} .
\end{eqnarray}
\begin{definition}
\label{def-lift}
For a given $\rho \in \Herpos$,
let $\Phi \in \pi_m^{-1}(\rho)$ be chosen arbitrarily. 
Then, for any $\Xi \in T_{\rho} \Herpos$, there exists the
unique tangent vector, $\ell_{\Phi}(\Xi) \in T_{\Phi}\inv$,
at $\Phi$ subject to
\begin{eqnarray}
\label{defeq-lift}
(\pi_{m*})_{\Phi}(\ell_{\Phi}(\Xi))=\Xi
\quad \mbox{and} \quad
\ell_{\Phi}(\Xi) \in \ker ((\pi_{m*})_{\Phi}).
\end{eqnarray}
The tangent vector $\ell_{\Phi}(\Xi)$ at $\Phi \in \inv$ is
called the horizontal lift of $\Xi \in T_{\rho}\Herpos$ to
$T_{\Phi}\inv$.
\end{definition}
The horizontal lift $\ell_{\Phi}(\Xi)$ is shown to take
the following form \cite{Rm3}.
\begin{proposition}
\label{prop-express-lift}
Let $\rho \in \Herpos$ be expressed in the form
\begin{eqnarray}
\label{diag}
\rho = h \Delta h^{\dagger}
\,\, \mbox{with} \,\, h \in U(m) , \,
\Delta = \mbox{\rm{diag}} (\delta_1 , \cdots , \delta_m) 
\, \mbox{s.t.} \,
\sum_{j=1}^{m}\delta_j=1 , \, \delta_j >0 
\, (j=1,\cdots ,m),
\end{eqnarray}
so that $\Phi \in \pi_m^{-1}(\rho)$ admits the
singular-value decomposition (\ref{sing-decomp}) with
$\lambda_j=\sqrt{\delta_j}$ ($j=1,\cdots,m$).
The horizontal lift, $\ell_{\Phi}(\Xi)$, of
$\Xi \in T_{\rho}\Herpos$ takes the form
\begin{eqnarray}
\label{express-lift}
\ell_{\Phi}(\Xi)
=
\frac{\sqrt{m}}{2} g 
\left(
\begin{array}{c} \Delta^{-1/2} 
(h^{\dagger} \Xi h 
  + \alpha_{\sqrt{\Delta}}(h^{\dagger} \Xi h) )
\\
O_{2^n-m, m}
\end{array}
\right) h^{\dagger} ,
\end{eqnarray}
where $\sqrt{\Delta}$ stands for the square root of the
diagonal matrix $\Delta$ and $\alpha_{\sqrt{\Delta}}$ is
given by (\ref{alpha}) with $\Lambda = \sqrt{\Delta}$.
\end{proposition}
In terms of the horizontal lift and the metric
$(\, \cdot \, , \, \cdot \,)$ of $\inv$, the Riemannian
metric of $\inv$ that we are seeking is defined as follows.
\begin{definition}
\label{Riemann-sub}
The Riemannian metric, $\Rmet{\cdot}{\cdot}$, that makes
$\pi_m : \inv \rightarrow \Herpos$ the Riemannian
submersion is defined to provide the inner product subject
to
\begin{equation}
\label{def-Riemann}
\Rmet{\Xi}{\Xi^{\prime}}_{\rho}
=
(\ell_{\Phi}(\Xi) , \ell_{\Phi}(\Xi^{\prime} ))_{\Phi}
\qquad
(\Xi, \Xi^{\prime} \in T_{\rho} \Herpos , \,
\rho \in \Herpos),
\end{equation}
where $\Phi \in \pi_{m}^{-1} (\rho)$ is chosen arbitrarily.
\end{definition}
Note that by the validity of (\ref{def-Riemann}) the
$\pi_m$ is meant to be the Riemannian submersion from
$(\inv, \, (\cdot , \cdot))$ to
$(\Herpos , \Rmet{\cdot}{\cdot})$. A straightforward
calculation shows the following.
\begin{proposition}
\label{expression-R}
Let $\rho \in \Herpos$ is expressed in the form (\ref{diag}).
The Riemannian metric $\Rmet{\cdot}{\cdot}$ of $\Herpos$ 
making $\pi_m$ the Riemannian submersion takes the form,
\begin{eqnarray}
\label{exp-Riemann}
\Rmet{\Xi}{\Xi^{\prime}}_{\rho}
=
\frac{1}{2} \sum_{j,k=1}^{m} 
\frac{1}{\delta_j + \delta_k}
	\overline{\chi}_{jk} \chi^{\prime}_{jk}
\quad \mbox{with} \quad
	\chi = h^{\dagger} \Xi h , \,\, 
	\chi^{\prime}= h^{\dagger} \Xi^{\prime} h.
\end{eqnarray}
\end{proposition}
\subsection{The quantum information geometry}
In the previous subsections, the space of positive
semi-definite Hermitean matrices, $\Herposemi$ is given
as SRCMQ through the dimension-reduction of $\Mnmone$ by
the $U(2^n)$ action. Since $\Herposemi$ can be looked upon
mathematically as the space of $m \times m$ density
matrices, it is quite natural for us to come to study
quantum information geometry of $\Herposemi$ or of its
open-dense subset $\Herpos$ representing SR${}^2$CMQ.  
In the following, the SLD quantum Fisher metric is studied
as a typical object, which is known to play a core role
in quantum statistical theory on the space of density
matirces \cite{FH}.
\par
We start with defining the symmetric logarithmic derivative
(SLD). An account for the SLD is given below together with
its classical counterpart: In classical theory, the Fisher
metric (namely, the Fisher information matrix) is utilized
to \lq measure' the distance between a pair of points in
the space of certain parametric probability distributions.
The logarithmic derivative of distribution functions is
defined to be the derivation of the likelihood functions for
those distributions, which plays a crucial role in defining
the classical Fisher metric. Accordingly, a quantum
counterpart of the logarithmic derivation is required to
define the quantum Fisher metric, which is however not
determined uniquely due to the non-commutativity among
matrices. As a conventional one, we introduce the symmetric
logarithmic derivative (SLD) as follows
\cite{FH} .
\begin{definition}
\label{def-SLD}
Let $\Xi$ be any tangent vector at
$\rho \in \Herpos$. The SLD, denoted by
${\cal L}_{\rho}(\Xi)$, for $\Xi$ is defined to be the
linear map that satisfies
\begin{equation}
\label{SLD}
\frac{1}{2}
\left\{
\rho {\cal L}_{\rho}(\Xi) + {\cal L}_{\rho}(\Xi ) \, \rho
\right\} 
=
\Xi  \qquad (\Xi \in T_{\rho}\Herpos).
\end{equation}
\end{definition}
A straightforward calculation shows the following:
\begin{proposition}
\label{expression-SLD}
Let $\rho \in \Herpos$ is expressed in the form
(\ref{diag}). With the notations
\begin{eqnarray}
\label{note-SLD-Xi}
\chi_{jk}=(h^{\dagger} \Xi h)_{jk},
\quad
L_{jk}=(h^{\dagger} {\cal L}_{\rho}(\Xi) h)_{jk}
\qquad (j,k=1,\cdots,m),
\end{eqnarray}
the SLD, ${\cal L}_{\rho}(\Xi)$, for
$\Xi \in T_{\rho}\Herpos$ takes the form
\begin{eqnarray}
\label{SLD-exp}
L_{jk} = \frac{2}{\delta_j + \delta_k} \chi_{jk}
\qquad (j,k=1,\cdots,m).
\end{eqnarray}
\end{proposition}
Note that the SLD cannot be extended to the whole
$\Herposemi$ due to (\ref{SLD-exp}) and to the fact that
any $\rho \in \partial \Herposemi$ admit the
null-eigenvalue. The SLD quantum Fisher metric is defined
as follows.
\begin{definition}
\label{def-Fisher}
The quantum SLD Fisher metric $\QFmet{\cdot}{\cdot}$
is defined to give the inner product
$\QFmet{\cdot}{\cdot}_{\rho}$ in $T_{\rho}\Herpos$
subject to
\begin{equation}
\label{defeq-Fisher}
\QFmet{\Xi}{\Xi^{\prime}}_{\rho}
=
\frac{1}{2} \Re \, \Big[
\trace \left[ \rho
            ( L_{\rho}(\Xi) L_{\rho}(\Xi^{\prime})
            +
              L_{\rho}(\Xi^{\prime}) L_{\rho}(\Xi))
       \right] 
\Big] 
\qquad
(\Xi , \Xi^{\prime} \in T_{\rho} \Herpos ).
\end{equation}
\end{definition}
Equations (\ref{SLD-exp}) and (\ref{defeq-Fisher}) are
combined together to yield an explicit expression of
$\QFmet{\Xi}{\Xi^{\prime}}_{\rho}$, which is compared
with (\ref{exp-Riemann}) to show the first main theorem of
this paper.
\begin{theorem}[The 1st main theorem]
\label{MAIN_THEOREM_1}
The quantum SLD Fisher metric $\QFmet{\cdot}{\cdot}$
of the space, SR${}^2$CMQ, of regular
relative-configurations of multi-qubit states in ordered
tuples coincides with the Riemannian metric
$\Rmet{\cdot}{\cdot}$ up to the multiplication constant $4$;
\begin{eqnarray}
\label{coincidence}
\QFmet{\Xi}{\Xi^{\prime}}_{\rho}
=
4 \Rmet{\Xi}{\Xi^{\prime}}_{\rho}
\qquad
(\Xi, \Xi^{\prime} \in T_{\rho}\Herpos , \, \rho \in \Herpos).
\end{eqnarray}
\end{theorem}
\par
Thus the pair of geometric structures of the SR${}^2$CMQ are
shown to coincide with each other up to the multiplication
constant. One is the Riemannian metric arising from the
geometry of the fibered space structure over SR${}^2$CMQ.
The other is the SLD quantum Fisher metric
arising from the (mathematical) identification of
SR${}^2$CMQ with the space of regular density matrices
\cite{Rm4}. The coincidence seems to be
accidental in the present circumstances though, since those
metrics have been given rise from different concepts and
then shown their coincidence by calculations. It would be
interesting to seek a good account for connecting those objects.
\section{An integrable gradient system on SR${}^2$CMQ}
This section is devoted to present an integrable gradient
system on the SR${}^2$CMQ, which has quantum information
features.
\subsection{The von-Neumann entropy on the SR${}^2$CMQ}
On regarding $\Herpos$ (SR${}^2$CMQ) as an open-dense
subspace of the space of density matrices, one of the
most typical quantum-information objects on $\Herpos$
would be the von-Neumann entropy defined by
\begin{eqnarray}
\label{vN}
S(\rho)
=
- \trace (\rho \log \rho ) \qquad (\rho \in \Herpos),
\end{eqnarray}
which can be extended to $\Herposemi$. On solving the
maximum-value problem of $S(\rho)$, the target state
$W \in \Mnmone$ for our search is characterized as follows.
\begin{proposition}
\label{target-max}
Any target tuple $W$ given by (\ref{W}) for the search is
projected through $\pi_m : \inv \rightarrow \Herpos$ to the
unique maximum point, $\displaystyle{\frac{1}{m}} I$, of
$S(\rho)$, where $I$ denotes the $m \times m$ identity
matrix.
\end{proposition}
Due to Prop.~\ref{target-max}, the projection,
$\{ \pi_m((-I_A \circ I_W)^k (W)) \}_{k=0, 1, 2, \cdots}$,
of the sequence generated by the search algorithm is
understood to get close to the maximum point $(1/m)I$, of
$S(\rho)$ if $k$ gets close to $(\pi \sqrt{2^n}-1)/2$.
This fact inspires us to study the gradient dynamical system
associated with $-S(\rho)$, all of whose trajectories
besides the stationary one tend to $(1/m)I = \pi_m (W)$.
\subsection{The equation of motion for the gradient system}
As shown in the previous section, we have endowed the pair
of non-Euclidean metrics to $\Herpos$, so that the gradient
operator has to be associated with either of those.  Since a
quantum information aspect is one of our interest on this
system, the SLD quantum Fisher metric $\QFmet{\cdot}{\cdot}$
is chosen to describe the gradient operator \cite{Rm5}.
Then the gradient vector of $-S(\rho)$ at $\rho$, denoted by
$(\mbox{grad}_{\rho} (-S))(\rho)$, is defined to satisfy
\begin{eqnarray}
\label{def-grad}
\QFmet{\, (\mbox{grad}_{\rho} (-S))(\rho)}{\, \Xi \,}_{\rho}
=
d(-S)_{\rho}(\Xi) \qquad (\Xi \in T_{\rho}\Herpos) ,
\end{eqnarray}
where $d(-S)_{\rho}$ on the rhs of (\ref{def-grad}) denotes
the exterior differential of $-S(\rho)$ at $\rho$.
In a more intuitive way, the rhs can be written as the
differentiation,
\begin{eqnarray}
\label{d(-S)}
d(-S)_{\rho}(\Xi)
=
\left. \frac{d}{d\tau} \right\vert_{\tau =0} 
\big[ -S(\gamma(\tau )) \big]
\end{eqnarray}
along a smooth curve $\gamma (\tau)$ on $\Herpos$ subject
to $\gamma (0)=\rho$ and $(d\gamma/d\tau)(0)=\Xi$
($0 \leq \vert \tau \vert < {}^{\exists}\varepsilon$).
The equation of motion for the gradient system on the
quantum information space $(\Herpos , \QFmet{\cdot}{\cdot})$
\cite{Rm4} with the potential $-S(\rho)$ is described by
\begin{eqnarray}
\label{grad-eq}
\dot{\rho} = - (\mbox{grad}_{\rho}(-S))(\rho) ,
\end{eqnarray}
where $\dot{\phantom{x}}$ indicates the differentiation
by the time-parameter $t$ henceforth.
\par
We are to express the rhs of (\ref{grad-eq}) more explicitly
through a geometric way \cite{UIH}. Let $\widetilde{S}(\Phi)$
be the $U(2^n)$ invariant function
\begin{eqnarray}
\label{wt-S}
\widetilde{S}(\Phi)
=
-\trace \Big[ \frac{1}{m} \Phi^{\dagger}\Phi \,
	 \log  (\frac{1}{m} \Phi^{\dagger}\Phi) \Big]
\end{eqnarray}
which satisfies $S \circ \pi_m = \widetilde{S}$.
Then the identities,
\begin{eqnarray}
\label{ident}
dS_{\rho}(\Xi)= d\widetilde{S}_{\Phi}(\ell_{\Phi}(\Xi))
\,\, \, \mbox{and} \, \,\,
\QFmet{\, \dot{\rho}}{\, \Xi \,}_{\rho}
=
4(\, \ell_{\Phi}(\dot{\rho}), \, \ell_{\Phi}(\Xi) \, )_{\Phi}
\quad
\left(
\begin{array}{l} \Xi \in T_{\rho}\Herpos , \\ \Phi \in \inv
\end{array}
\right) ,
\end{eqnarray}
are put together to provide the explicit expression,
\begin{eqnarray}
\label{exp-eq-grad}
\dot{\rho} 
=
\rho \Big[
\big( \trace (\rho \log \rho )\big) I - \log \rho 
\Big],
\end{eqnarray}
of (\ref{grad-eq}), where $\ell_{\Phi}$ denotes the
horizontal-lift operation (\ref{defeq-lift}) and $I$ the
$m \times m$ identity matrix. The $\log \rho$ is the
logarithm of the matrix $\rho$, which is well-defined for
any $\rho \in \Herpos$. 
\subsection{Solution and integrability}
We are to solve (\ref{exp-eq-grad}) under the ansatz
\begin{eqnarray}
\label{ansatz-rho}
\rho(t)= h(t) \Delta (t) h(t)^{\dagger}
\quad \mbox{with} \quad 
\Delta(t)=\mbox{diag} (\delta_1(t), \cdots , \delta_m(t)),
\, h(t) \in U(m),
\end{eqnarray}
together with
\begin{eqnarray}
\label{h-const}
h(t) \equiv h(0) \in U(m) \qquad (t \in {\bf R}).
\end{eqnarray}
Under (\ref{ansatz-rho}) and (\ref{h-const}),
Eq.~(\ref{exp-eq-grad}) reduces to the system of
differential equations,
\begin{eqnarray}
\label{DE-delta}
\dot{\delta_j}
=
\delta_j(t) 
\left[ \sum_{k=1}^{m} \delta_k(t) \log \delta_k(t) \right]
- \delta_j(t) \log \delta_j(t)
\qquad (j-1,\cdots ,m),
\end{eqnarray}
for $\delta_j$s, whose initial value problem is solved to be
\begin{eqnarray}
\label{sol-delta}
\delta_j (t)
=
\left( \sum_{k=1}^{m} 
\left( \delta_k(0) \right)^{\exp(-t)} \right)^{-1}
\left( \delta_j(0) \right)^{\exp(-t)}
\qquad (j=1, \cdots ,m).
\end{eqnarray}
Surprisingly, Eq.~(\ref{DE-delta}) takes exactly the same form
as the equation of motion for the gradient system on the
space of multinomial distributions with the negative entropy
as the potential \cite{N}. 
Combining (\ref{sol-delta}) with the ansatz (\ref{ansatz-rho}),
we have the following.
\begin{proposition}
\label{solution-grad}
Let the triple,
$(\Herpos , \QFmet{\cdot}{\cdot} , -S(\rho))$,
express the gradient system on the quantum information
space $(\Herpos , \QFmet{\cdot}{\cdot})$ with the potential
$-S(\rho)$, where $S(\rho)$ denotes the von-Neumann entropy
defined by (\ref{vN}). The solution of the equation of motion
(\ref{grad-eq}) for
$(\Herpos , \, \QFmet{\cdot}{\cdot} , \, -S(\rho))$ takes
the form,
\begin{eqnarray}
\label{sol-grad}
\rho (t)
=
h(0) \,
\left[ 
\trace \left[ \left(\Delta(0)\right)^{\exp (-t)} \right] 
\right]^{-1}
\left(\Delta(0)\right)^{\exp (-t)} \, h(0)^{\dagger} ,
\end{eqnarray}
where $h(0) \in U(m)$ and
$
\Delta(t)
=\mbox{\rm{diag}} (\delta_1(0), \cdots , \delta_m(0))
$
are given by the initial condition,
$\rho(0)= h(0) \Delta (0) h(0)^{\dagger}$ with
(\ref{ansatz-rho}).
\end{proposition}
\par
Now that we have got the solution of the gradient system
explicitly, we show the integrability of the gradient system
$(\Herpos , \, \QFmet{\cdot}{\cdot} , \, -S(\rho))$ in turn.
Note that the integrability here means the existence of
$\mbox{dim} \Herpos -1$ constants of motions being mutually
independent. We start with the construction of $m-2$ 
independent constants of motion in $\delta_j$s
($j=1,\cdots,m$).
From (\ref{sol-delta}) and (\ref{sol-grad}),
we find
\begin{eqnarray}
\label{invariants}
\frac{\log \delta_{j+2}(t) - \log \delta_{j+1}(t) }
     {\log \delta_{j+1}- \log \delta_{j}(t) }
\qquad (j=1,\cdots,m-2)
\end{eqnarray}
as constants of motion being mutually independent.
Further from Proposition~\ref{solution-grad} with the
expression (\ref{ansatz-rho}), we find the unitary matrix
$h(t)$ is kept invariant to be $h(0)$, from which we take
$m(m-3)/2$ independent constants of motion \cite{Rm6}.
We have thus found mutually-independent
$\mbox{dim} \Herpos -1$ (=$m(m+1)/2-2$)
constants of motion for the gradient system. To summarize,
we reach to the second main theorem of this paper.
\begin{theorem}[The 2nd main theorem]
\label{MAIN_THEOREM_2}
The gradient system on the quantum information space
$(\Herpos , \QFmet{\cdot}{\cdot}$ with the negative von
Neumann entropy $-S(\rho)$ as the potential is integrable
in the sense that it admits mutually-independent
$\mbox{dim} \Herpos -1$ constants of motion.
\end{theorem}
\section{Concluding remarks}
We have made the geometric studies on SR${}^2$CMQ, the
space of regular relative-configurations of multi-qubit
states in ordered tuples which comes from the space
for the quantum search for an ordered tuple of
multi-qubit states. We have shown that SR${}^2$CMQ admits
the Riemannian metric arising from the fibered space
structure over SR${}^2$CMQ and the SLD quantum Fisher
metric, which turn out to be the same up to the multiplication
constant.
\par
On the quantum information space
$(\Herpos, \QFmet{\cdot}{\cdot})$ \cite{Rm4},
the gradient system with the negative von-Neumann
entropy $-S(\rho)$ is taken to study, which is shown to
be integrable. The equation (\ref{DE-delta}) describing
the time-evolution of spectra $\delta_j$s of $\rho$
turns out to be the same as that of the gradient system
associated with multinomial distributions
\cite{N} appearing in the classical statistical theory,
although we have to take the additional degree-of-freedom
generated by $U(m)$ into account. Further, our integrable
gradient system might be of new type. To study more on our
gradient system, the Lax representation would be worth
being investigated.
\par
On closing this paper, the authors wish to mention of
a future investigation: The integrability of our gradient
system has inspired us to think of a possibility of a
search algorithm with a kind of \lq convergence' through
a discretization of the gradient equation.
\begin{acknowledgments}
All the authors would like to thank Prof.~T.~Iwai and
Prof.~Y.~Nakamura at Kyoto University for their valuable
comments on this work.
\end{acknowledgments}


\begin{thebibliography}{999}
\bibitem{Rm1}
Although our method looks a naive generalization of Miyake
and Wadati \cite{MW} to the ordered tuple case, they differ from
each other in a big extent: Our geometric method is base on
the left action of a unitary group leaving the \lq relative
configuration' of multi-qubit states in tuples, while such
a group action cannot be thought of in \cite{MW}.
\bibitem{Rm2}
The case of $\Phi$ with a naught column vector will not be
dealt with, since all such $\Phi$s will be excluded in
thinking of SR${}^2$CMQ.
\bibitem{Rm3}
The singular-value decomposition (\ref{sing-decomp}) 
is utilized very effectively in the geometric studies
on SRCMQ and SR${}^2$CMQ \cite{UHI,UIH}, although the
usage is not apparently given due to the length of this
paper.
\bibitem{Rm4}
Although SR${}^2$CMQ does not seem to have a physical
meaning in view of its construction, it admits the same
mathematical structure as the space of regular density
matrices. With this observation in mind, we wish to
refer to $(\Herpos , \QFmet{\cdot}{\cdot} )$ as the quantum
information space also henceforth.
\bibitem{Rm5}
We can also consider the gradient system endowed with the
Riemannian metric $\Rmet{\cdot}{\cdot}$. Due to
Theorem~\ref{MAIN_THEOREM_1}, the gradient systems with
$\QFmet{\cdot}{\cdot}$ and with $\Rmet{\cdot}{\cdot}$ turn
out to be equivalent up to the constant multiple in
time-parameters.
\bibitem{Rm6}
The replacement of $h$ by
$h \, \mbox{diag}(\varepsilon_1, \cdots , \varepsilon_m)$
(
$\vert \varepsilon_1 \vert = \cdots 
 = \vert \varepsilon_m \vert=1$
) provides the same $\rho$ in (\ref{diag}). Hence the
degree-of-freedom in charge of $h \in U(m)$ in (\ref{diag})
is $\mbox{dim}U(m)-m = m(m-3)/2$.
\end{thebibliography}
\end{document}